\documentclass[%
reprint,
superscriptaddress,
 amsmath,amssymb,
 aps,
]{revtex4-2}
\setcitestyle{authoryear,round}
\usepackage{graphicx}
\usepackage{bm}
\usepackage{color} 
\usepackage[hidelinks]{hyperref}

\usepackage{fancyhdr} 
\begin{document}

\title{A global simulation of the dynamo, zonal jets and vortices on Saturn}

\author{Rakesh Kumar Yadav}
\email{To whom correspondence should be addressed: rakesh\_yadav@fas.harvard.edu}
\affiliation{Department of Earth and Planetary Sciences, Harvard University,  Cambridge, MA 02138, USA}

\author{Hao Cao}
\email{hcao@epss.ucla.edu}
\affiliation{Department of Earth and Planetary Sciences, Harvard University,  Cambridge, MA 02138, USA}
\affiliation{Department of Earth, Planetary, and Space Sciences, University of California, Los Angeles, CA 90095, USA}

\author{Jeremy Bloxham}
\email{jeremy\_bloxham@harvard.edu}
\affiliation{Department of Earth and Planetary Sciences, Harvard University,  Cambridge, MA 02138, USA}

\date{\today}

\begin{abstract}
{The fluid dynamics in planet Saturn gives rise to alternating east-west jet streams, large cyclonic and anticyclonic vortices, and a dipole-dominant magnetic field which is highly axisymmetric about the planetary rotation axis. Modelling these features in a self-consistent manner is crucial for understanding the dynamics of Saturn's interior and atmosphere. Here we report a turbulent high-resolution dynamo simulation in a spherical shell which produces these features {\em simultaneously} for the first time. A crucial model ingredient is a long-hypothesised stably stratified layer (SSL), sandwiched between a deep metallic hydrogen layer and an outer low-conductivity molecular layer, born out of limited solubility of Helium inside metallic Hydrogen at certain depths. The model spontaneously produces polar cyclones and significant low and mid latitude jet stream activities in the molecular layer. The off-equatorial low-latitude jet streams partially penetrate into the SSL and interact with the magnetic field. This helps to axisymmetrize the magnetic field about the rotation axis and convert some of the poloidal magnetic field to toroidal field, which appears as two global magnetic energy rings surrounding the deeper dynamo region. The simulation also mimics a distinctive dip in the fifth spherical harmonic in Saturn's magnetic energy spectrum as inferred from the Cassini Grand Finale measurements. Our model highlights the role of an SSL in shaping the fluid dynamical and magnetic features of giant planets, as exemplified at Saturn.}
\end{abstract}

\maketitle

\section{Introduction}

The gas giant planet Saturn, well known for its spectacularly bright icy rings, has many more defining properties. On its visible surface, the different colored gases and clouds sketch the profiles of the planet-circulating east-west winds which alternate in direction  from the equator to the poles and reach speeds of several 100s of meters per second \citep{lavega1982, lavega2000}. One of these zonal jets takes the form of a regular hexagon in the north polar region \citep{godfrey1988}. Embedded among these jets are large vortical storms which can rotate either in the clockwise or anticlockwise direction \citep{trammell2014,trammell2016}. A polar cyclonic vortex is found at each geographic pole \citep{lavega2006}. In addition, the angle between Saturn's spin axis and the axis of the dominant magnetic dipole ($<$0.007$^{\circ}$) is almost 3 orders of magnitude smaller than that of the Earth \citep{dougherty2018,cao2020}. Furthermore, its non-dipolar magnetic field is also highly axisymmetric about the rotation axis, much more so than any other object in the solar system \citep{dougherty2018,cao2020}. 

These properties of Saturn are a direct consequence of the rotating fluid dynamics occurring in its interior and atmosphere, and provide a great opportunity for building and constraining theoretical models of Saturn's interior fluid dynamics. Through magnetohydrodynamic interactions, the electrical conductivity of a fluid inside a gas giant planet like Saturn will likely act as an important factor in demarcating regions with different flow properties \citep{stevenson2003}. It has been hypothesised \citep{liu2008} that due to the varying electrical conductivity of liquid hydrogen inside Saturn (and Jupiter) \citep{nellis2000, french2012}, the magnetohydrodynamic interactions largely separate Saturn's interior into an outer strong zonal flow region and an inner one with much reduced zonal flows. The detailed mechanism involved in this separation is an active area of research \citep{christensen2020,gastine2021}. 

A significant body of work has been dedicated to understanding how the zonal jets and vortices are created in the outer layer of Saturn. Broadly speaking, there are two interpretations. In one, which we may call a "bottom up" approach see \cite{kulowski2021} for a recent discussion of this issue), the zonal jets are considered to be surface manifestations of deep (thousands of kilometers)  concentric, axially-aligned, rotating cylinders generated through the convective, rotating turbulence. This idea was put forth by \citet{busse1976} and has been extensively explored in many numerical simulations studies \citep[e.g.,][]{christensen2001, aurnou2001, heimpel2005, kaspi2009, jones2009, gastine2014, yadav2020}. In the other approach, the zonal jets are thought to exist in a thin atmospheric layer (100s of kilometers or less), similar to the Earth's atmosphere, within which the shallow water approximation is typically employed \citep{williams1978,cho1996a,cho1996b,lian2010,liu2010, cabanes2020}. Here, the driving sources are commonly assumed to be latent heat released through cloud condensation and solar irradiation. 

The gravity data collected by the Cassini Grand Finale has shown that Saturn's gravity perturbations can be explained by slightly modified surface zonal winds penetrating to nearly 9000 km \citep{galanti2019}. Therefore, the zonal jets are very deep indeed. However, teasing out the jet-driving mechanism remains difficult because even a shallow production mechanism confining in the atmosphere of Saturn could also lead to very deep jets \citep{showman2006,lian2008}. 

Saturn's highly axisymmetric magnetic field is exceptional in light of the Cowling's theorem \citep{cowling1933} which states that a dynamo in an incompressible fluid cannot maintain a purely axisymmetric magnetic field. Although, several of the original requirements have been relaxed in subsequent work \citep{hide1982,kaiser2018}, we can safely assume that a dipole dominant magnetic field perfectly aligned with the spin axis cannot be generated by a turbulent dynamo itself. We can extend this line of thought to assume that even the magnetic fields like the one existing in Saturn, where the dipole tilt angle is extremely small, likely require special mechanisms for becoming highly axisymmetric.

\cite{stevenson1979} theoretically modelled the interaction of Helium and Hydrogen at physical conditions relevant to giant planets \citep[also see][]{salpeter1973}. He proposed that Helium concentration may be at or close to saturation as hydrogen transitions from molecular to metallic with increasing depths. This would lead to differentiation (`rain out') of Helium and the development of a stably stratified layer in the outer parts of the metallic hydrogen layers \citep{salpeter1973,  stevenson1982, stevenson1982axi, hubbard1985}. Recent {\em ab initio} studies suggest immiscibility of Helium in hydrogen for pressures lower than 1 Mbar at Saturn-like conditions \citep[e.g., ~see][]{lorenzen2009, morales2013}, which is around 65\% of Saturn's radius. However, a good estimate of the  thickness of such a stable layer remains highly uncertain \citep{stevenson1977, fortney2011}. Using such a layer as an ingredient, a mechanism was proposed to axisymmetrize a planetary dynamo: ``{\em If a conducting fluid shell is undergoing spin-axisymmetric differential rotation and overlies the dynamo generating region of a planet then it is capable of greatly reducing the non-spin-axisymmetric components of the generated field}'' \citep{stevenson1982axi}.

This idea has been tested by several subsequent models. A kinematic dynamo study showed that an SSL can help axisymmetrize the magnetic field, but not always \citep{love2000}. Fully non-linear dynamo simulations performed with an overlying SSL showed that the SSL was very effective at removing non-axisymemtric small-scale magnetic fields \citep{christensen2008}. Another study imposed latitudinal heat-flux variations to excite axisymmetric flow inside an SSL on top of a dynamo region and showed that the flows inside the SSL can help either axisymmetrize or destabalize the magnetic field depending on their properties \citep{stanley2008,stanley2010}. In the most recent study \citep{yan2021} in this context, a thick SSL with an imposed latitudinally-varying heat-flux perturbation is used to promote Saturn-like dipole dominant magnetic fields with very small dipole tilt angles of about 0.066$^{\circ}$ which is still about one order of magnitude larger than the latest observational upper bound \citep{cao2020}.

A few studies have investigated models without an SSL (and hence also without any latitudinally-varying boundary heat-flux) in which only the extremely steep  drop in the electrical conductivity in the outer layers of Saturn (and Jupiter) is considered. It is generally found that strong zonal flows are excited (in the equatorial region) in the outer layer where the electrical conductivity is lower \citep{duarte2013,dietrich2018}. The dipole dominant solutions, a requirement for modelling Saturn's dynamo, are less stable in these setups. However, we recently showed \citep{yadav2022}  that such setups without an SSL can produce dipole-dominant solutions with exceptionally small dipole tilts of $\approx$ 0.0008$^{\circ}$ if the non-linearities in the simulation (promoted by low convective supercriticalities) are small enough. The axial quadrupole in that model, however, is many orders of magnitude smaller than that observed at Saturn. 

To summarize, the zonal jets on Saturn are typically modelled separately from the dynamo, while the interior dynamo is modelled either with an SSL or with an overlying low-conductivity layer. In this study we try to consolidate these approaches and present a global model where we simulate the deep dynamo, an SSL, and the molecular hydrogen layer on top. It must be noted, however, that our "global" model still lacks the meteorological physics (e.g. cloud formation, rapid density variation, and solar insolation) occurring in the outer few hundred kilometers of Saturn. The model details given below will make it clear that our main aim is to investigate the non-linear interactions between different dynamically important regions in giant planets -- specifically the interaction of an SSL with other regions. We do not attempt to match the exact background state of an interior model proposed for Saturn. To date, the state of Saturn's interior is an active area of research; for recent developments see \citet{iess2019, militzer2019, movshovitz2020, dewberry2021, mankovich2021}.

\section{Method}
\subsection{{\em MHD Equations}}
For our simulations, we use the same tool and methodology as used by \citet{gastine2021} where they investigate the effect of a stably stratified layer on the fluid dynamics and the resulting dynamo mechanism in the context of Jupiter. We refer the reader to their study for a detailed description of the methodology employed to model a sandwiched SSL; also see \citet{dietrich2018SSL}. Here we only highlight some of the essential details. 

To simulate some of the magnetohydrodynamical processes operating inside a giant planet, we first assume the model planet has a perfectly spherical shape and is bounded by an inner boundary at $r_i$ and an outer boundary at $r_o$. The spherical shell rotates about a fixed spin axis ($\hat{z}$) with angular velocity $\Omega$. The motions of the fluid inside the spherical shell are modelled using the LBR (Lantz-Braginsky-Roberts) version of the anelastic approximation \citep{braginsky1995, lantz1999}. This approximation assumes that a thermodynamic quantity, for instance, $x(r,\theta,\phi,t)$, is a combination of a background state, $\tilde{x}(r)$, and a perturbation, $x'(r,\theta,\phi,t)$. We work with a non-dimensional system of variables and use the shell thickness $d=r_o-r_i$ as the length scale, and the viscous diffusion time $d^2/\nu$, where $\nu$ is the kinematic viscosity, as the time scale. The magnetic field, $\vec{B}$, is normalized by $\sqrt{\tilde{\rho_o}\mu_o\lambda_i\Omega}$, where $\tilde{\rho_o}$ is density on the outer boundary, $\mu_o$ is magnetic permeability, and $\lambda_i$ is magnetic diffusivity on the inner boundary. Note that the square of the magnetic field magnitude in this unit is known as the Elsasser number. The entropy, $s$, and the fluid density, $\rho$, is scaled by their values at the outer boundary. 

The following system of equations govern the time evolution of velocity ($\vec{u}$), entropy, and magnetic field:
\begin{gather}
\nabla\cdot(\tilde{\rho}{\vec u})=0, \\
\left(\frac{\partial \vec{u}}{\partial t}+\vec{u}\cdot\vec{\nabla}\vec{u}\right)
= -\vec{\nabla}{\frac{p'}{\tilde\rho}} 
- \frac{2}{E}\hat{z}\times\vec{u}
- \frac{Ra}{Pr}\tilde{\alpha}\tilde{T}\vec{g} \,s'\ \nonumber \\
+\frac{1}{Pm_i\,E \,\tilde{\rho}}\left(\vec{\nabla}\times \vec{B}\right)\times \vec{B}
+ \frac{1}{\tilde{\rho}} \vec{\nabla}\cdot \mathsf{S}, \label{eq:vel}   \\
\tilde{\rho}\tilde{T}\left(\frac{\partial s'}{\partial t} +
\vec{u}\cdot\vec{\nabla} s' +u_r\frac{d\tilde{s}}{dr} \right) =
\frac{1}{Pr}\vec{\nabla}\cdot\left(\tilde{\rho}\tilde{T}\vec{\nabla} s'\right) +\nonumber \\
\frac{Pr\,Di}{Ra}\Phi_\nu + 
\frac{Pr\,Di\,\lambda_{norm}}{Pm_i^2\,E\,Ra}\left(\vec{\nabla}
\times\vec{B}\right)^2, \label{eq:entropy}\\
\vec{\nabla}\cdot \vec{B}=0, \\
\frac{\partial \vec{B}}{\partial t} = \vec{\nabla} \times \left( \vec{u}\times\vec{B}\right)-\frac{1}{Pm_i}\vec{\nabla}\times\left(\lambda_{norm}\vec{\nabla}\times\vec{B}\right). \label{eq:mag}
\end{gather}

In the above equations, $p$ is pressure, $T$ is temperature, $\tilde{\alpha}$ is the dimensionless expansion coefficient, $\vec{g}$ is gravity vector (gravity magnitude increases linearly with radius), $\lambda_{norm}$ is the radially-varying magnetic diffusivity normalized by its value at the inner boundary. The traceless rate-of-strain tensor $\mathsf{S}$ in Eqn.~2 is given by
\begin{gather}
S_{ij}=2\tilde{\rho}\left(e_{ij}-\frac{1}{3}\delta_{ij}\vec{\nabla}\cdot\vec{u}\right) \text{with}\,\,
e_{ij}=\frac{1}{2}\left(\frac{\partial u_{i}}{\partial x_{j}}+\frac{\partial u_{j}}{\partial x_{i}}\right) \nonumber
\end{gather}
where $\delta_{ij}$ is the identity matrix. The viscous heating contribution in Eqn.~3 is given by
\begin{gather}
\Phi_{\nu}=2\tilde{\rho}\left[ e_{ij} e_{ji} - \frac{1}{3} (\vec{\nabla}\cdot\vec{u})^2 \right]. \nonumber
\end{gather}
The non-dimensionalization process introduces the following fundamental control parameters in the equations: 
\begin{gather}
E=\frac{\nu}{\Omega\,d^2}; \,\,\, Ra=\frac{\alpha_o\,T_o\,g_o\,d^3\Delta s}{c_p\nu\kappa} \nonumber \\ 
Di=\frac{\alpha_o\,g_o\,d}{c_p}; \,\,\, Pr = \frac{\nu}{\kappa}; \,\,\, Pm_i=\frac{\nu}{\lambda_i} \nonumber
\end{gather}
where subscript `o' denotes the values at the outer boundary. 

\subsection{{\em Model Setup}}
Unlike the study of Gastine \& Wicht \citep{gastine2021} where an interior model of Jupiter was taken as a strong guiding factor, here we only model interactions among some features which we consider important for the interior dynamics of Saturn-like planets. Therefore, our study is an exploratory one. The primary aim being to study how a thick SSL might affect the workings of the planetary dynamo and the deep atmosphere. The background profile of the thermal variables is given in Supplementary Figure 1.

We set the inner boundary of the shell at $r_i=0.27r_o$. The convection in the fluid shell is driven by an imposed entropy contrast. Within the spherical shell, we assume that there is a stably stratified layer between 0.62$r_o$ and 0.8$r_o$ which resists large-scale overturning convection and provides a restoring force to radial fluid motions. This is achieved by imposing a constant positive radial gradient (a free parameter in the model) in the background entropy profile within the SSL and a value of -1 elsewhere in the shell. The imposed entropy radial gradients in the convective layer and in the SSL are joined smoothly with each other using hyperbolic tangent functions. We refer the reader to \citep{gastine2021} for more details of this implementation strategy \citep[also see][]{gastine2020, takehiro2001}. An important parameter in this context is the ratio of the Brunt-V{\"a}is{\"a}l{\"a} frequency and the rotation rate ($\Omega$). With the control parameters chosen in the simulation, this ratio is about 1.5 in the SSL. In the study of \citet{gastine2021}, this ratio is about 10. The lower value utilized in our setup allows for a greater penetration of convection into the SSL \citep[e.g., see][]{takehiro2001}. 

The viscosity and the thermal conductivity of the fluid are assumed constant throughout the spherical shell. The electrical conductivity, however, has a strong radial dependence: it stays constant from $r_i$ to about 0.78$r_o$ and exponentially decreases by more than five orders of magnitude within a thickness of about 0.1$r_o$; outside of 0.9$r_o$, the fluid is assumed to be electrically insulating. 

With this model setup, there effectively exist four dynamically important regions which self-consistently interact with each other (see Fig.~\ref{fig1}a): a convective dynamo region (from 0.27$r_o$ to 0.62$r_o$), an electrically conducting SSL (0.62$r_o$ to 0.8$r_o$), a `semi' conducting and convective region from 0.8$r_o$ to 0.9$r_o$, and a purely hydrodynamic and convective region (from 0.9$r_o$ to $r_o$).

\subsection{{\em Control Parameters and Boundary Conditions}} The choice of the control parameters is largely driven by the available computing resources, a need for lowering the fluid viscosity as much a possible (quantified by the Ekman number) and pushing the (magnetic) Reynolds number to high values. Given these constraints, the main control parameters we could model are: $Ek=7\times10^{-7}$, $Ra=2.2\times10^{11}$, $Pr=1$, $Pm=0.4$, $Di=6$. The mechanical boundaries at $r_i$ and $r_o$ are impenetrable and stress free. For the magnetic field, the lower boundary at $r_i$ acts as a conductor having the same electrical conductivity as the fluid in the dynamo region; the radial level $0.9r_o$ and higher act as an electrical insulator for the magnetic field. The entropy is assumed constant on each boundary at $r_i$ and $r_o$ with a prescribed entropy contrast between the two.

Due to the complexities involved in running such a demanding low-viscosity simulation, several compromises need to made: (1) It is not possible to run such a simulation with a small seed magnetic field and let the dynamo build up the large scale field since it will take impractically large computing power. Therefor, we start the simulation from a case having a saturated dynamo state. (2) The Ekman number needs to be decreased in incremental stages to reach $7\times10^{-7}$. In the reported simulation here, we started with a simulation at $Ek=1.3\times10^{-6}$ and decreased the Ekman number to $7\times10^{-7}$ in stages in the first $\approx$15\% of the total time evolution. (3) The Rayleigh number was increased to $2.2\times10^{11}$ from $4\times10^{10}$ in the first $\approx$50\% of the time evolution. (4) The $Pr$ was constant throughout the simulation but the $Pm$ was lowered from 0.6 to 0.4 in the first $\approx$50\% of the time evolution. (5) Due to the low viscosity associated with $Ek=7\times10^{-7}$, the flow exists on very small length scales which were not possible to be resolved given the computing grid we used. Therefore, a numerical technique called hyperdiffusion was used throughout the simulation where flows on length scales smaller than a set value are suppressed more than those with larger length scales. This is a commonly utilized strategy in low Ekman number simulations \citep{kuang1999,soderlund2012,gastine2021,heimpel2022}. The utilized hyperdiffusion was such that the Ekman number gradually increases by about 100 after spherical harmonic (SH) degree $\sim$70 in the first half and after SH degree $\sim$100 in the second half of the simulation. 

The simulation could only be evolved for about 0.0045 viscous diffusion time, or about 0.011 magnetic diffusion time for the parameters mentioned above. It translates to about 1000 rotations of the spherical shell. The relatively short time span of the simulation, measured by the diffusion time, is an unavoidable drawback at such extreme parameters.  Therefore, the statistical steadiness of the simulation is not guaranteed. The short time evolution of the simulation still consumed about 5.5 million core hours which is a rather large amount of computing time. Despite these limitations, we believe the richness of the results obtained warrants a documentation and a discussion. 

\subsection{{\em Simulation Code and Grid Size}}
The Eqns.~1 through 5, operating under the above control parameters and boundary conditions, are solved using the open source code {\tt MagIC} (\href{https://magic-sph.github.io/}{https://magic-sph.github.io}) \citep{gastine2012}. It uses the pseudo-spectral approach and projects the primitive variables in the latitudinal-longitudinal directions onto the spherical harmonic functions  while the radial component is projected onto the Chebyshev polynomials. It also uses the toroidal-poloidal decomposition to maintain strict divergenceless condition for the relevant quantities. MagIC utilizes the open source library SHTns \citep{shtns} to performe the spherical harmonic transformations. The grid size, given as number of grid points in longitudinal, latitudinal, and radial, respectively is [2112, 1056, 480] for the first 90\% of the simulation evolution and increases to [2560, 1280, 640] in the final 10\%.

\begin{figure*}[!ht]
\centering
\includegraphics[width=0.8\linewidth]{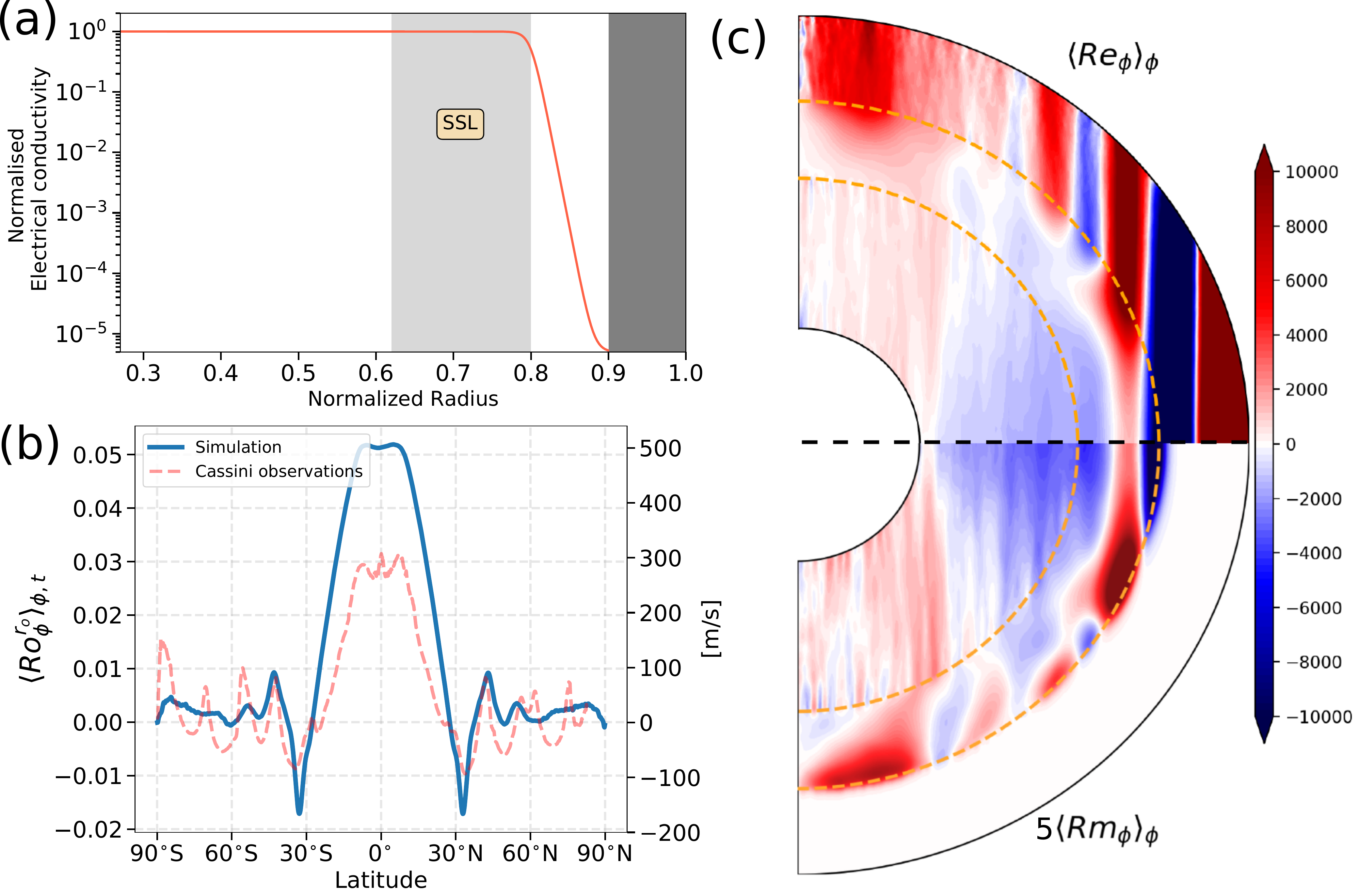}
\caption{Panel {\bf a} shows the variation of the electrical conductivity as a function of the radius. The light shaded region (from radius 0.62 to 0.8) is the stably stratified layer and the dark shaded region (radius 0.9 to 1) has zero electrical conductivity. Panel {\bf b} shows the axisymmetric zonal flow on the surface of the simulation and on Saturn's cloud deck, given in terms of the Rossby number $u/(\Omega r_o)$ on the left axis and in m/s on the right axis. We use the Cassini dataset (from the `CB' filter) covering a period from year 2004 to year 2009 presented by \cite{garcia2011}. The data has been rescaled to a recent estimate of Saturn's rotation period $10^{h}33^{m}38^{s}$ \citep{mankovich2019}. We assumed a mean radius of $5.8232\times10^7$m for Saturn to convert the observed wind speeds to a radius-based Rossby number. The top half of Panel {\bf c} shows the instantaneous axisymmetric zonal flow on a meridional plane, given in terms of the Reynolds number $Re$=$u\,d/\nu$; the lower half of the panel shows 5 times the magnetic Reynolds number $Rm$ defined by $u\,d/\lambda$, where $\lambda$ is the local magnetic diffusivity. The factor of 5 is applied for the $Rm$ half panel to share the same colormap with the $Re$ half panel.}
\label{fig1}
\end{figure*}

\section{Results and Discussion}
In this section we present our results. We begin by looking at the properties of the surface and internal zonal flow. We then describe some properties of large vortices that are produced near the simulation surface. In the later half of the section, we present the properties of the magnetic field produced by the dynamo region. We end the section with a comparison of the magnetic field morphology with the one observed by Cassini at Saturn.

\subsection{Surface zonal flow} 
We begin by first describing the flow properties arising in the simulation.  In Fig.~\ref{fig1}b we plot the instantaneous, axisymmetric zonal flow on the outer boundary of the simulation. As a reference we also plot the corresponding values obtained for Saturn from cloud tracking \citep{lavega2000,vasavada2006,lavega2006,read2009}. Both data are given in terms of the Rossby number, $Ro=u/(\Omega r_o)$, on the left axis where the length scale is the planetary radius and in m/s on the right axis. The peak strength of the simulation zonal flow in the equatorial region is larger than the Saturn's value by about 65\%. The latitudinal profile of the broad, eastward equatorial zonal jet is similar to Saturn's equatorial jet. The flows in the first westward jet in the simulation are faster as well (by about a factor of two) and peak slightly closer to the equator than the similar jet on Saturn. Prior studies \citep{christensen2001, aurnou2001, heimpel2005, jones2009, gastine2014,yadav2020} investigating the origin of zonal flows in deep spherical shells show that the location of the first westward zonal jet is largely governed by the location of the lower boundary of the spherical shell, with deeper shells promoting westward jets at higher latitudes. In our simulation, the location of the top of the SSL acts as the lower barrier (or a pseudo `boundary') for the outer hydrodynamic layer. Therefore, one may expect that if the SSL was located slightly deeper, then the westward jet would shift to higher latitudes and would be more in agreement with the Saturn's values; however, it should be mentioned that a deeper shell depth might also promote a stronger or weaker equatorial jet.


The two off-equatorial eastward jets peaking at around 45$^{\circ}$ north and south are reduced in strength by a factor of about five compared to the equatorial jet, similar to the observations in Saturn's northern hemisphere (Fig.~\ref{fig1}b). At latitudes polewards of about 45$^{\circ}$, longitudinally coherent zonal jets are not present in our simulations, although a longitudinal averaging does produce a tiny eastward peak at around 55$^{\circ}$. For latitudes within about 15$^{\circ}$ of both rotational poles, we see some zonal jet activity with a weak, but broad and coherent eastward zonal flow in both hemispheres (discussed later; see Fig.~\ref{fig3}). Note that, on the other hand, Saturn's cloud deck features coherent zonal jets at all latitudes.

\subsection{Zonal flow in the deep interior} 
In Fig.~\ref{fig1}c we present the variation of the zonal flow as a function of depth. Since the fluid shell is rotating along a fixed axis and has a low Ekman number, we can expect the features to be highly influenced by the planetary spin in accordance with the Proudman-Taylor theorem \citep{proudman1916,taylor1923}. Indeed, in the hydrodynamic layer ($r >0.8 r_o$) overlying the SSL, the zonal flow (Fig.~\ref{fig1}b) in the mid to low latitudes largely projects downwards along spin-aligned cylinders. At higher latitudes, the zonal flow lacks such cylindrical coherence. The strong spin-axis aligned zonal jets in such conditions are thought to be maintained by the Reynolds stresses in rotating, turbulent, convective flows \citep[e.g.,][]{busse1994,gastine2013}. 

\begin{figure}[!ht]
\centering
\includegraphics[width=1\linewidth]{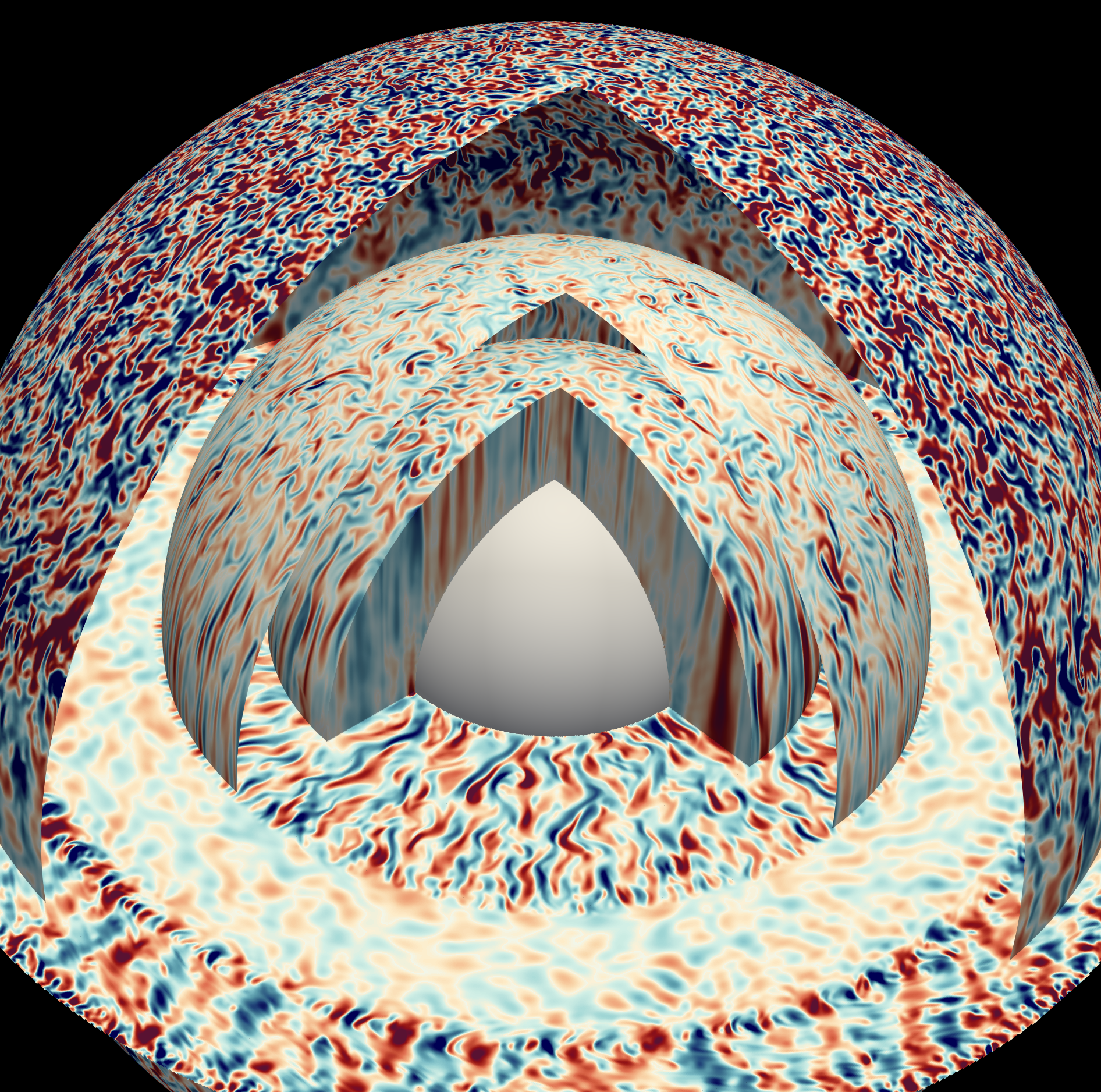}
\caption{Radial velocity (given in Reynolds number) on the equator and at various depths; the color map is saturated at $\pm$5000 and the red (blue) shades correspond to +ve (-ve) values.}
\label{fig2}
\end{figure}

\subsection{Comparison of the zonal flow with earlier studies} 
As we briefly alluded to in the introduction, zonal jet formation has been extensively investigated with non-magnetic convection in rapidly rotating spherical shells. It has been shown that the flow boundary conditions on the spherical shell boundaries play a deciding role. For models where both the bottom and the top boundaries are stress free (or free slip), conherent zonal jets with great strength are readily excited at all latitude when the convective flow reaches sufficiently high Reynolds numbers \citep{christensen2001, aurnou2001,heimpel2005,jones2009,gastine2014,yadav2020,yadav2022}. If both boundaries are assumed to be no-slip (or rigid), then it is much harder to excite coherent zonal jets \citep[e.g., see][]{yadav2016}. If the boundary conditions are mixed, i.e. no-slip condition at the bottom and free-slip condition at the top, then only one zonal jet outside the tangent cylinder (a spin-aligned cylinder touching the inner boundary) is formed \citep{aurnou2004,jones2009,heimpel2022}. The quenching of zonal flow in the presence of rigid boundaries is largely due to the enhanced friction provided by the Ekman boundary layers \citep[e.g., see][]{jones2007}. In the absence of Ekman boundary layers, the zonal jets can saturate at much larger strengths, which is only limited by the much lower volume friction \citep{christensen2002,jones2007}. This also helps explain the case for mixed boundaries: any local cylindrical section of the fluid domain \textit{outside} the tangent cylinder only experiences free-slip boundaries on its northern and southern ends, which allows strong zonal flows. For flows inside the tangent cylinder, the bottom no-slip boundary provides enough friction to damp coherent zonal flows.

\begin{figure*}[!ht]
\centering
\includegraphics[width=1\linewidth]{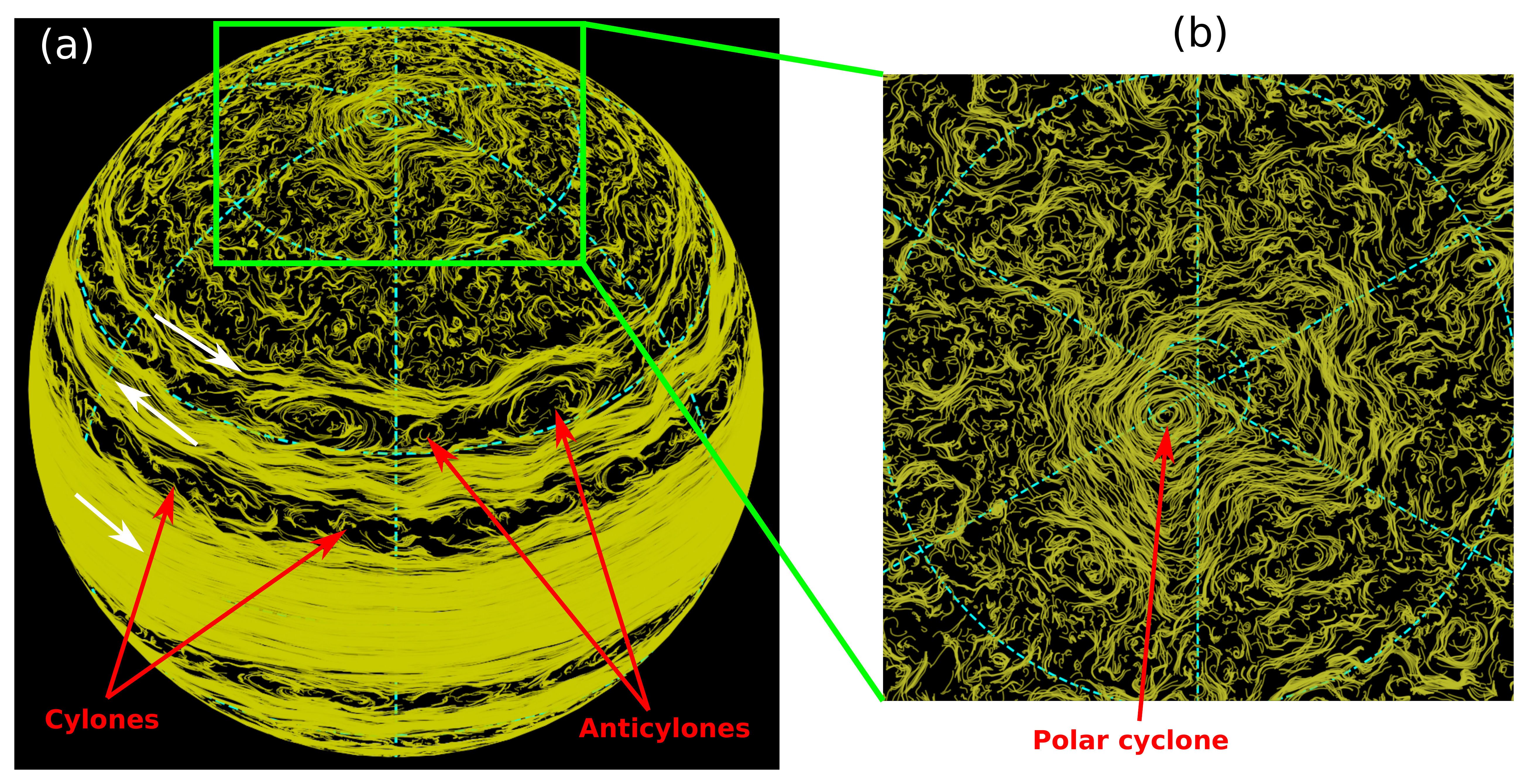}
\caption{Panel {\bf a} shows the instantaneous streamlines highlighting the zonal jets and (anti)cyclonic vortices at around 0.93$r_o$.  The white arrows show the general rotational direction of the equatorial and the two northern coherent zonal jets. The streamlines in the north polar region are highlighted in Panel {\bf b}. An animation of both panels, showing the evolution (about 12 rotations) of the streamlines as the background flow evolves, is provided in the supplementary material. The plots were generated using the OceanParcels package \citep{parcels} which allows tracking of massless particles using a background advection velocity field.}
\label{fig3}
\end{figure*}

\begin{figure*}[!ht]
\centering
\includegraphics[width=0.9\linewidth]{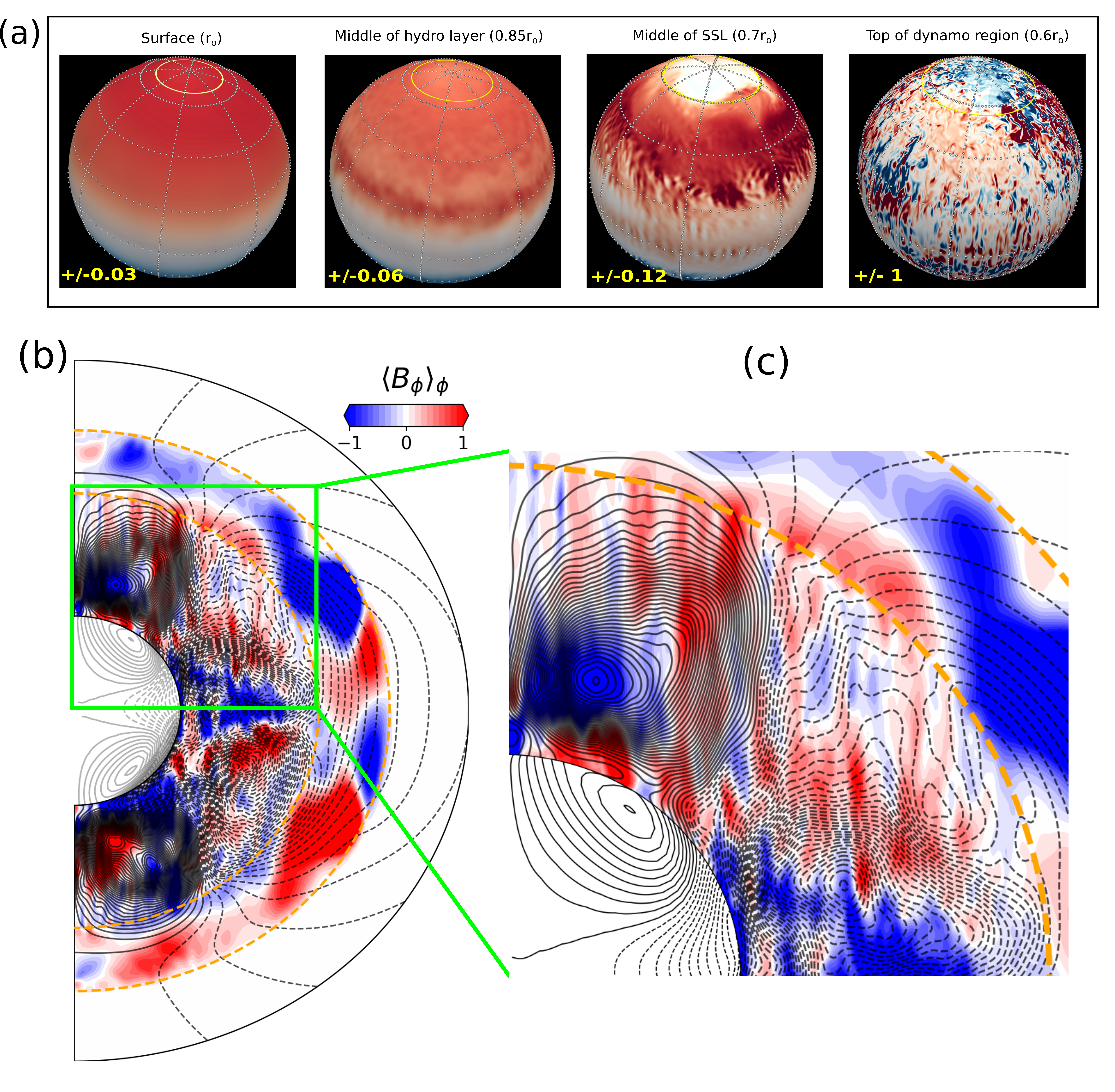}
\caption{Panel {\bf a} shows the magnitude of the radial magnetic field at various representative depths. The field magnitude is given in Elsasser number units; see Method section for definition. The range of the color map is shown at the lower left side of each sub-panel. The intersection of the tangent cylinder created by the inner core at $r_i$ and a given spherical surface at various radii is shown using a yellow circle. Panel {\bf b} shows the axisymmetric zonal magnetic field as color map and the axisymmetric poloidal magnetic field as contour lines; solid (dashed) lines represent (anti)clockwise field lines. A zoomed in section, showing the dense field lines in the northern hemisphere of the dynamo region is shown in Panel {\bf c}.}
\label{fig4}
\end{figure*}

The zonal jet features mentioned for the mixed boundary conditions have also been found in some dynamo models where electrically conducting fluid in the deeper part of the shell (`dynamo' region) and low conductivity fluid in the shallower layer (`hydro' region) are simulated simultaneously. If these dynamo simulations produce strong dipole-dominant magnetic fields, then the Lorentz force effectively renders the dynamo region into a rigidly rotating shell which provides a drag `boundary condition' (mimicking some aspects of a no-slip boundary condition) for the fluid in the shallower low-conductivity region on top. A magnetic tangent cylinder is setup \citep{dietrich2018} at the interface of the dynamo and the hydrodynamic region which determines the width of the broad, eastward equatorial jet \citep{duarte2013,gastine2014Jup,yadav2022}.

In our simulation, some of the zonal jets present in the outer convective shell manage to penetrate significantly into the stably stratified layer (Fig.~\ref{fig1}c). Remarkably, the strong westward jet flanking the equatorial jet manages to pierce through the SSL almost cylindrically. The adjacent higher-latitude, eastward jets in the north and in the south penetrate to a lesser degree, are less cylindrical in the SSL, and join with a much weaker flow at the equator. Note that since the SSL provides only a radial restoring force, it can indeed be expected that  zonal flow may penetrate into the SSL. The recent study of \citet{gastine2021} shows much less penetration of zonal flows into the SSL. We attribute this difference to the less `stiff' SSL with lower B-V frequency used in our study (see {\em Model Setup} subsection).   

The convective dynamo region ($r <0.62 r_o$) also features zonal flows which are much weaker in strength but are maintained steadily. Such zonal flows with a broad westward drift in the mid to low latitudes and an eastward drift in the higher latitudes are commonly produced in the geodynamo simulations and are maintained by a thermal wind balance \citep{aubert2005}. The weak westward flow from the dynamo region penetrates into the lower half of the SSL and is contained in the mid to lower latitudes. Note that a weak westward flow is also present in the dynamo region of \citet{gastine2021}.

\subsection{Radial flow features} 
The radial flows present in the simulation are shown in Fig.~\ref{fig2}. The SSL featuring much weaker radial flow (with much fainter colors) is immediately evident. The radial flows within the SSL are predominantly internal gravity waves (IGWs). The flows in the inner dynamo region, showcasing spin-aligned, radially-stretched convective plumes, appear very much like those seen in low Ekman number geodynamo simulations \citep[e.g., see][]{yadav2016}.

\subsection{Zonal jet truncation} 
Using a detailed force balance analysis, \citet{gastine2021} showed that the primary mechanism that limits the strongest zonal flows to the outer low conductivity layers on top of the SSL is similar to the one outlined by the theoretical model of \citet{christensen2020}. As the zonal winds reach fluid with sufficiently high conductivity, they result in, via induction, Lorentz forces that drive a meridional circulation pattern. In the SSL, this circulation modifies the background stratification and the entropy profile, promoting a thermal wind balance which acts to truncate the zonal winds. Given the broad similarities between our model setup and the results we obtain with the one in \citet{gastine2021}, it is likely that the zonal flow truncation follows a similar mechanism.

\subsection{Large cyclones and anticyclones} 
Before we move on to describe the properties of the dynamo generated magnetic field, we highlight the rich dynamics present in the outer layers of the simulation. Figure \ref{fig3}a shows the equatorial jet, and the westward and eastward jets (present in both hemispheres) at higher latitudes. The edges of these jets are cyclonic and anticylonic shear regions which promote and sustain strings of cyclones and anticyclones. We recently demonstrated such a mechanism of vortex generation in spherical shell with convectively driven flows \citep{yadav2022}. This mechanism is known to exists on both Saturn and Jupiter \citep[e.g., see][]{vasavada2005}. Most of these large scale vortices in the simulation are deep, spin-aligned structures existing in the low conductivity region on top of the SSL. Although we do not have a general overview of how deep are the vortices on giant planets, a recent orbit of Juno on Jupiter was close enough to show that at least some of the vortices could be 100s of kilometers deep \citep{bolton2021}. 

Note that the current simulation lacks the preferential production of giant anticyclones as reported in our earlier study \citep{yadav2022}. As we hypothesized in that study, large plumes from a deep dynamo region are needed to excite anticyclonic activity in the overlying hydrodynamic layer. In the simulation we report here, such a mechanism is not possible since the thick SSL on top of the dynamo region does not allow any plume to pass through it. This may help explain why, Jupiter, which may only have a thin SSL \citep[e.g. see][]{wahl2017,debras2019}, possibly allowing some plumes to go through, has an anticyclonic great red spot and a high preference for anticyclones, while Saturn lacks such preference.

The region from approximately 45$^{\circ}$ -- 75$^{\circ}$ north (south) contain no coherent jet activity and the vortices that form do not have a preferred rotation sense (see the animation of the Fig.~\ref{fig3} provided in the SI). In the polar region (Fig.~\ref{fig3}b), however, a coherent cyclone is present with an accompanying serpentine eastward jet (similarly in the south). The polar cyclone is not stationary and meanders about the pole. Such polar cyclones were recently reported in convectively driven spherical shell models \citep{garcia2020,yadav2020,heimpel2022}. The serpentine jet visible in the simulation's polar region is somewhat reminiscent of the polygonal jet, shaped as a hexagon, on Saturn \citep{godfrey1988}. However, it lacks the remarkable coherence seen on Saturn and in our previous study \citep{yadav2020} where we reported polygonal jets in a convectively driven thin spherical shell.


\begin{figure*}[!ht]
\includegraphics[width=\textwidth]{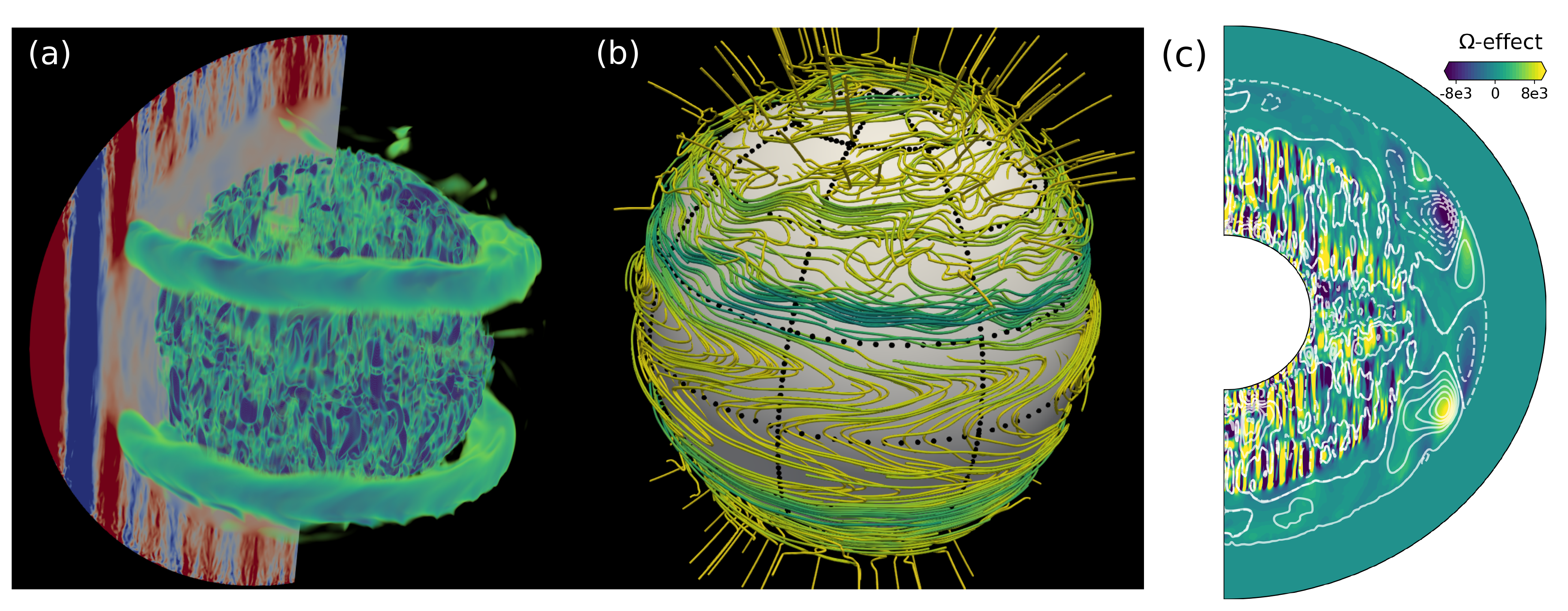}
\caption{ Panel {\bf a} shows the local zonal flow as color map on a meridional section and the magnetic energy as volumetric rendering where the field magnitude is assigned different opacities depending on their magnitudes, rendering low magnitude regions transparent or translucent. This technique helps to accentuate important energetic regions. Panel {\bf b} shows some of the magnetic field lines for radii larger than 0.7$r_o$. A fly-through animation of Panel {\bf a} and {\bf b} is provided in the supplementary material. Panel {\bf c} shows the axisymmetric toroidal magnetic field as contour lines and the $\Omega$-effect (see main text for definition) as color map.} \label{fig5}
\end{figure*}

\subsection{Dipole dominant magnetic field} 
Let us now turn to the properties of the dynamo generated magnetic field in the simulation. The radial component of the magnetic field at various depths is shown in Fig.~\ref{fig4}a. On the surface of the simulation, the magnetic field has very smooth character and is highly dipole-dominant (carrying about 95\% of the magnetic energy). At a depth in the lower half of the outer molecular layer, some small scale features start appearing due to the increasing electrical conductivity of the fluid (Fig.~\ref{fig1}a). The magnetic field is still dipole-dominant. As we reach the middle of the SSL, where the electrical conductivity has saturated to its maximum value, large deviations in morphology compared to the surface develop; here, the dipole carries only about 2\% of the magnetic energy. The most noticeable deviation being the depletion of radial magnetic field inside the tangent cylinder to the inner boundary at $r_i$. At radial depths reaching the top of the deep dynamo, where electrical conductivity is high and convection is taking place, the magnetic field is dominated by small scale structure. The magnetic field has completely changed in character and the dipole mode, which was dominant near the surface, now carries only about 0.5\% of the magnetic energy. A breakdown of the poloidal magnetic energy across various length scales and at different radii is provided in the Fig.~\ref{fig7}. Remarkably, the radial magnetic field inside the tangent cylinder has reversed in direction compared to that at the surface. Such weakening of radial magnetic field inside the tangent cylinder has been seen in some geodynamo simulations \citep[see a recent investigation and discussion presented in][]{cao2018}.

\begin{figure*}[!ht]
\centering
\includegraphics[width=1\linewidth]{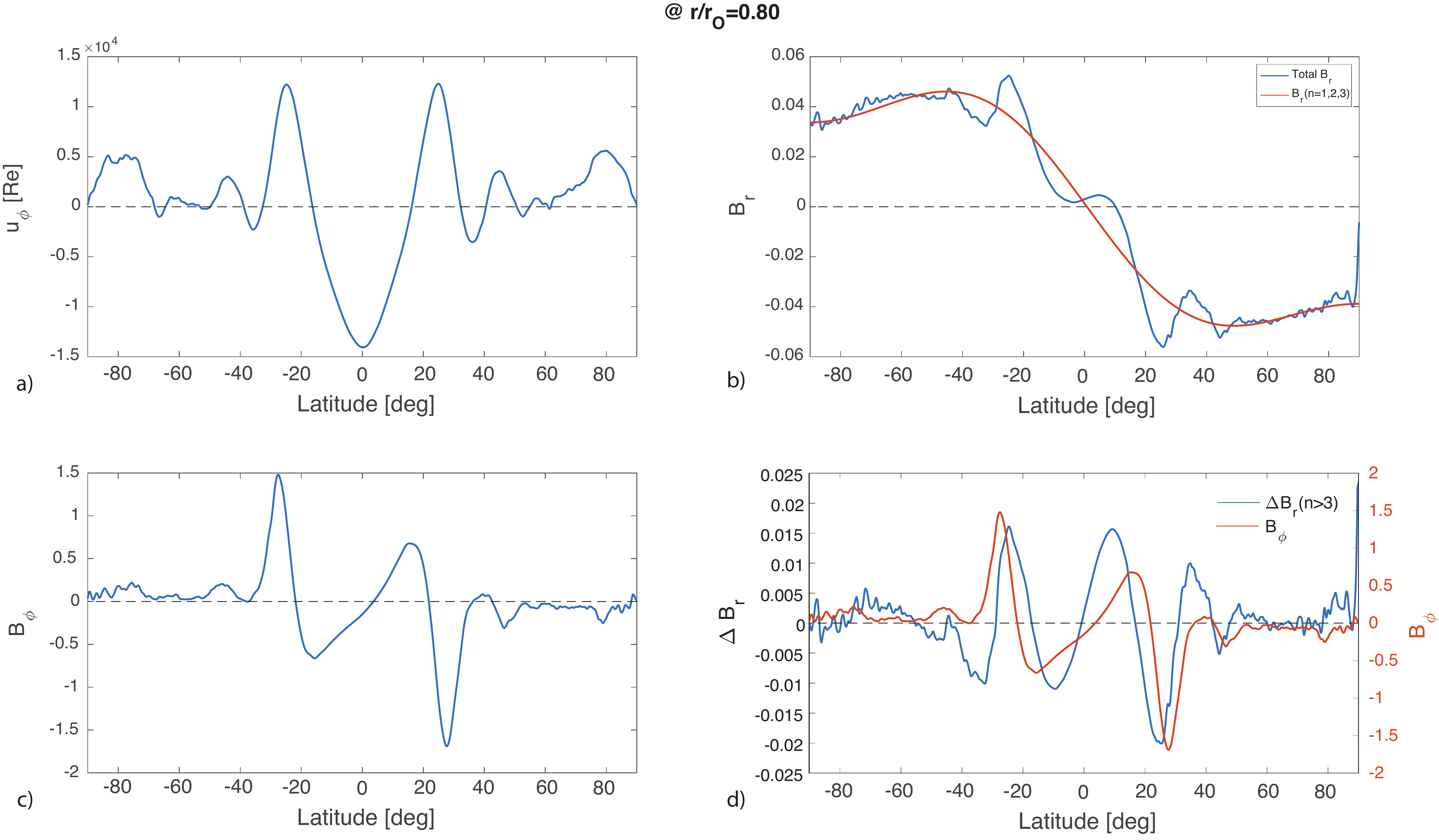}
\caption{Axisymmetric components of the zonal flow and magnetic field at the top of the SSL. Panel {\bf a} shows the zonal flow. Panel {\bf b} shows the total as well as the large scale component (only containing dipole, quadrupole, and octopole) of the radial magnetic field. Panel {\bf c} shows the azimuthal magnetic field. Panel {\bf d} compares the smaller scaled radial magnetic field with the total azimuthal magnetic field. All these quantities are at the top of the SSL which is at 0.8$r_0$.}
\label{fig6}
\end{figure*}

\begin{figure}[!ht]
\centering
\includegraphics[width=1\linewidth]{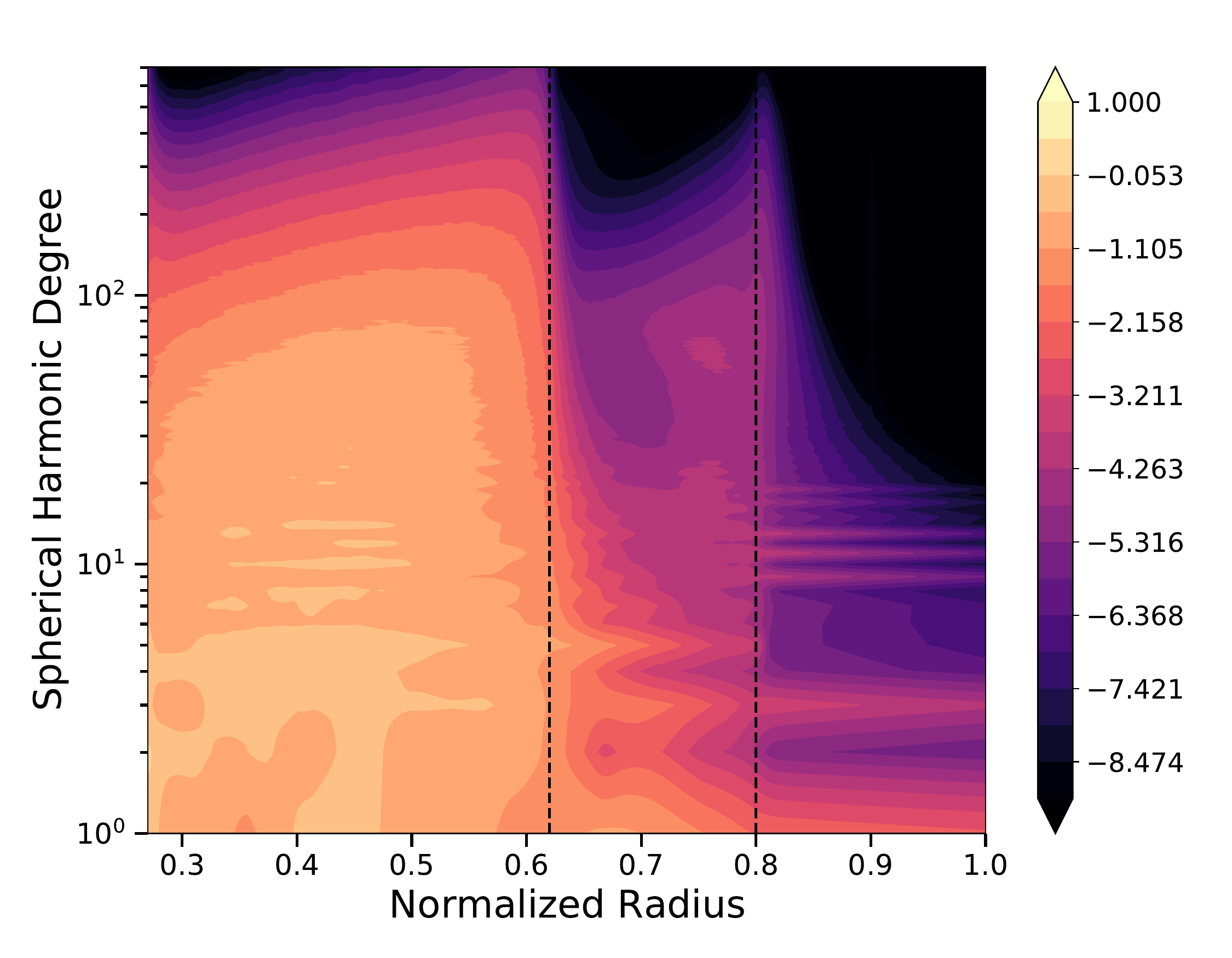}
\caption{Variation of the {\em logarithmic} poloidal magnetic energy, normalized by its overall maximum value, contained in different spherical harmonic degrees as we go from the innermost radius to the outermost radius. The boundaries of the stably stratified layer are shown using dashed vertical lines.}
\label{fig7}
\end{figure}

The axisymmetric toroidal and poloidal magnetic field shown in Fig.~\ref{fig4}b and \ref{fig4}c provide further insights into the magnetic field morphology. The poloidal field lines  demonstrate the transition from a complex field inside the dynamo region to a largely dipole-dominant one in the outer layers. It is evident that the reversal of radial magnetic field inside the tangent cylinder is a consequence of the poloidal field lines getting pulled back into the dynamo region as opposed to connecting to the other hemisphere. The counter rotating magnetic loop established inside the tangent cylinder is likely a consequence of the weak, but coherent, eastward zonal flow present inside the tangent cylinder and the associated meridional circulation \citep{cao2018}. 

\subsection{Rings of magnetic energy} 
In terms of the toroidal magnetic field shown in Fig.~\ref{fig4}b and \ref{fig4}c, the most remarkable features are found in the SSL. At low latitude, each hemisphere contains a strong toroidal magnetic field region, which is followed by another broader and much stronger one at higher latitudes. In Fig.~\ref{fig5}a we present an orthographic view to elucidate where the magnetic energy is concentrated. Not surprisingly, the deep dynamo region ($r< 0.62 r_o$) stands out as a central sphere carrying most of the magnetic energy. Moreover, two distinct magnetic energy rings hovering above the deep dynamo, one in north and another in south, are clearly visible. The field lines shown in Fig.~\ref{fig5}b clearly demonstrate that these magnetic energy rings are composed primarily of  toroidal field lines that are wrapped up in the azimuthal direction. Note that the two patches visible near the equator in Fig.~\ref{fig4}c are significantly weaker, and, therefor, are filtered out by the transparency function adopted in  Fig.~\ref{fig5}a. Azimuthally wrapped magnetic field structures were previously reported in a thin shell dynamo simulation of a Sun-like star \citep{brown2010}. There, such structures were not as azimuthally coherent and were part of the dynamo-wave process, unlike in our study where they form outside of the convective dynamo region and are relativley stable.

The local zonal flow shown on a meridional slice in Fig~\ref{fig5}a demonstrates that the location of the magnetic energy rings roughly coincides with the termination of the first off-equatorial eastward jet in both hemispheres. This gives us a hint that the interaction of the jets with the electrically conducting fluid in the SSL is likely producing strong  shear which generates the toroidal magnetic field through the classical $\Omega$-effect \citep{krause1980}. Such a mechanism essentially bends poloidal magnetic fields into toroidal morphology at the expense of the kinetic energy stored in the zonal flows. Following \citep{brown2011}, we quantify the $\Omega$-effect as:
\begin{gather}
\Omega = \bar{B_r}\frac{\partial}{\partial r}\left(\frac{\bar{u_{\phi}}}{r}\right) + \frac{\bar{B_{\theta}}\sin\theta}{r}\frac{\partial}{\partial\theta}\left(\frac{\bar{u_{\phi}}}{\sin\theta}\right),
\end{gather}
where the overbar denotes azimuthal averaging. In Fig.~\ref{fig5}c, we plot the axisymmetric component of $\lambda(r)\Omega$ on a meridional plane. In the deep dynamo layer, the contours of the toroidal magnetic field have no discernible correlation with the $\Omega$-effect color map. In the SSL, however, the color map almost perfectly aligns with the contours in the mid to low latitude regions. This strong correlation indicates that the magnetic energy rings present in the SSL are due to the $\Omega$-effect. Therefore, the production of the toroidal magnetic field in the SSL is intimately linked to the termination of the zonal jets in the SSL.

Another way to inspect the interaction of the zonal jets and magnetic field is to look at their behavior at a certain depth. In Figure \ref{fig6}, we plot the axisymmetric components of the zonal flow, the radial magnetic field, and the azimuthal magnetic field at the top of the SSL (0.8$r_o$). The profile of the radial magnetic field (Fig.~\ref{fig6}b) and the zonal flow (Fig.~\ref{fig6}a) at this depth motivates us to look at smaller scale fluctuation in the radial field. By removing the lowest three spherical harmonic degree contributions ($\ell=1,2,3$) from the total radial magnetic field, we separate the smaller scale fluctuations and compare it to the azimuthal magnetic field at the same depth (Fig.~\ref{fig6}d). It is immediately evident that there are clear correlations among these magnetic field components and the zonal flow; for instance, the first off-equatorial peaks are located around 25$^{\circ}$ north and south for zonal flow as well as the zonal magnetic field and the smaller scaled radial magnetic field. Similar correlations were envisioned by \citet{cao2017b}. Note that the peaks in the azimuthal magnetic field and the smaller scaled radial magnetic field are shifted by about a quarter of the phase. Such an interaction was modelled by \cite{cao2017b} using mean-field electrodynamics in the context of jet-magnetic field interaction at Jupiter and Saturn; also see \cite{galanti2021} for application of this idea for constraining the depth of the winds in the two gas giants.

\begin{figure}[!ht]
\centering
\includegraphics[width=1\linewidth]{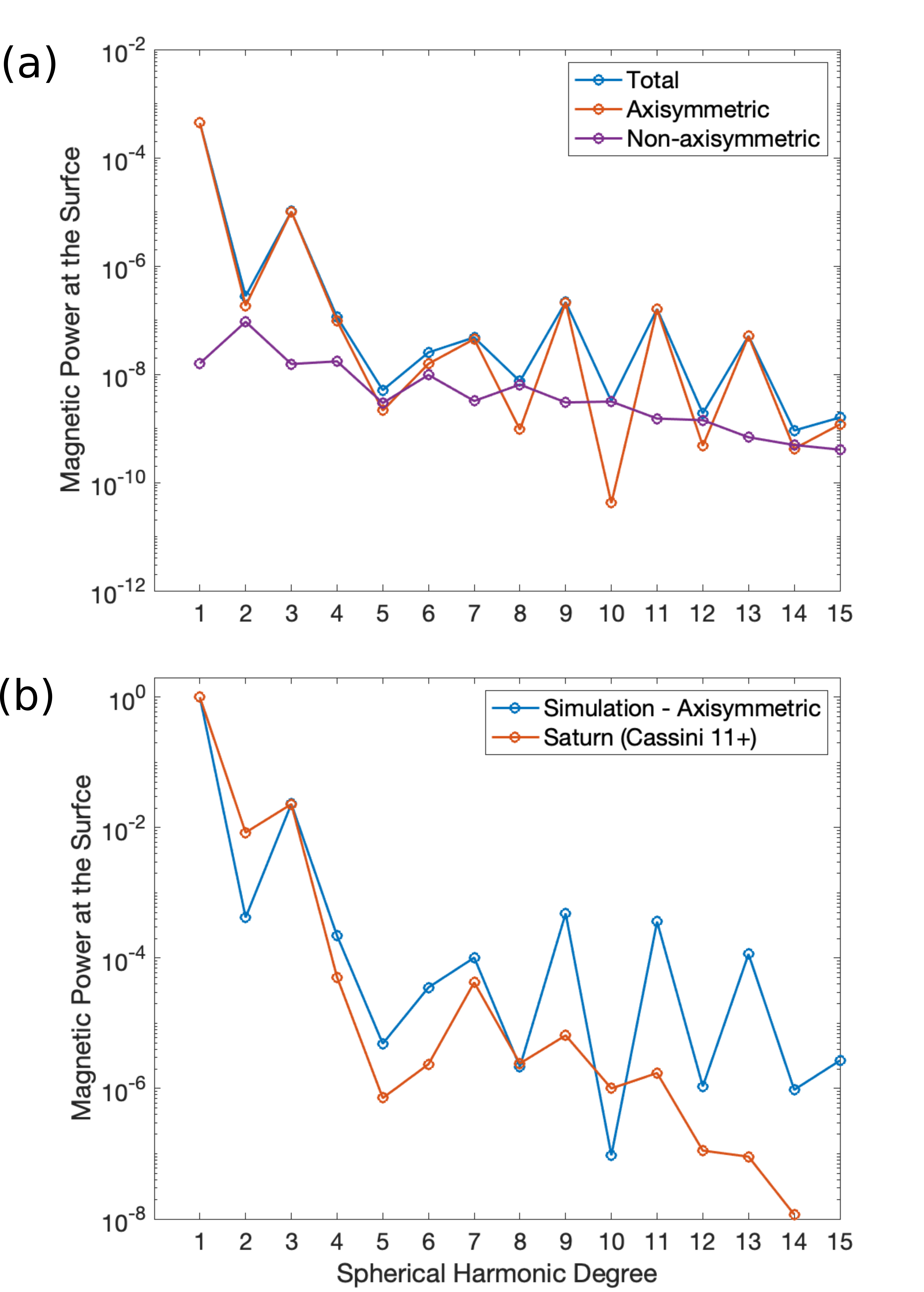}
\caption{Panel {\bf a} shows the (instantaneous) magnetic power contained in the first fifteen spherical harmonics on the simulation surface. Both axisymmetric and non-axisymmetric components are shown. Panel {\bf b} compares the axisymmetric power in the magnetic field with the values obtained for Saturn \citep{cao2020}. Note that both spectra are normalized by the power in the dipole mode.}
\label{fig8}
\end{figure}

\subsection{Comparing the magnetic field spectrum with Cassini observations} 
The magnetic power spectrum displayed in Fig.~\ref{fig8}a shows that the spherical harmonic degrees from 1 through 7 carry most of the energy in their axisymmetric component. For degrees eight and higher, the behavior is reversed. The magnetic field spectrum compares favorably (Fig.~\ref{fig8}b) with the one inferred for Saturn based on the Cassini Grand Finale magnetic field measurements \citep{dougherty2018,cao2020}. A good agreement with Saturn's magnetic field up to spherical harmonic degree eight, spanning almost six orders in magnitude of the magnetic power, is remarkable. Beyond degree eight, the simulation generally contains more magnetic energy at different length scales compared to Cassini observations. Note that in Fig.~\ref{fig8}b we have normalized the spectra so that the dipole power is at unity for comparison. Due to the mismatch in control parameters of the simulation and those at Saturn (for instance, much higher fluid viscosity in the simulation), it is not entirely straightforward to compare the magnetic field magnitudes.

We must note here that the spectrum of the simulation shown in Fig.~\ref{fig8} is simply a time shot and the similarity with Saturn's spectrum might not endure. As might be expected for a dynamical non-linear system, the spectrum may evolve into other states in which the characteristic dip at harmonic degree five goes away and a peak could appear instead; the latter is a common feature in dipole-dominant dynamo simulations \citep[e.g., see][]{duarte2013}. For Saturn, since we do not have a long record of its magnetic field morphology, its magnetic field spectrum might have been significantly different at different epochs.

\section{Summary \& Discussion}
The rich phenomenology of the fluid dynamics and magnetic field observed at Saturn has motivated many to propose novel driving mechanisms. To explain the exceptional level of axisymmetry in the Saturn's dynamo, \citet{stevenson1979} proposed the presence of a stably stratified layer (SSL) in which zonal flows may be able to reduce the non-axisymmetric components of a dynamo generated field deep within the planet. We set out to investigate how such a layer would impact the overall dynamics of the planet. Assuming a spherical shell geometry, we model a convective dynamo layer from 0.27 to 0.62, an electrically conducting and stably stratified layer form 0.62 to 0.8, and a convective, low-conductivity outermost layer from 0.8 to 1 (Fig.~\ref{fig1}a). 

The outermost layer self-consistently generates coherent, alternating zonal flows that reach the mid-latitudes (Fig.~\ref{fig1}b and \ref{fig1}c). These strong jets spawn large cyclonic or anticyclonic vortices depending on the local shear direction in the region where jets change direction (Fig.~\ref{fig2}a). In higher latitudes, the zonal jet activity is diminished and is replaced by (anti)cyclonic vortices with no clear preference (Fig.~\ref{fig2}b). A cyclonic polar storm exists on both rotation poles. We attribute this unprecedentedly rich dynamics to the low Ekman number ($7\times10^{-7}$) and high level of turbulence (Reynolds numbers reaching 5000 or higher) achieved in the simulation. The multiple zonal jets are largely confined to the low conductivity layer above the SSL. 

The deep dynamo generates a magnetic field with the octupole harmonic degree ($\ell=3$) mode being the dominant carrier of energy within the dynamo region.  However, after passing through the SSL and the low conductivity outer layers to the surface, the magnetic field morphology becomes strongly dipole dominant (Fig.~\ref{fig4} and Fig.~\ref{fig7}). Typically, the dipole mode is also tilted with a small angle of about 0.5 degrees with respect to the rotation axis. 

The eastward oriented zonal jet situated around midlatitudes on the surface, projects cylindrically downwards along the direction of the rotation axis and interacts with the magnetic field in the outer half of the SSL. This leads to an efficient local $\Omega$-effect where the poloidal magnetic field gets converted to toroidal morphology. The resultant toroidal magnetic field stands out in the magnetic energy space as two rings (one in each hemisphere) surrounding the deep dynamo region (Fig.~\ref{fig5}). 

At certain epochs in the simulation, the magnetic field spectrum on the surface compares favorably, for spherical harmonic degrees eight or less, with that observed at the present-day Saturn. On smaller length scales, the simulation produces more magnetic energy than the present-day Saturn (Fig.~\ref{fig8}).  

Although the features described above do not precisely match those observed at Saturn, the several qualitative similarities we obtain is made remarkable by the fact that they are all produced in one single simulation without any special care given to tune the results. Therefore, the idea of a thick SSL inside Saturn just above the deep dynamo appears to take us very close to many of Saturn's observed features. 

We can certainly imagine a few directions in which the current model can be improved. One way to `hide' magnetic energy from the surface is to convert the poloidal magnetic field into toroidal morphology which can not escape the outer non-conducting fluid layers. Our simulation lacks coherent zonal flow from mid to high latitudes. Since low Ekman numbers promote strong, coherent zonal flows \citep{cabanes2017}, we can expect a similar simulation at even lower Ekman number to produce coherent zonal jets at even higher latitudes. These zonal flows might help to convert more of the poloidal energy to toroidal energy at various length scales, helping to put the simulation more in line with the Saturn's internal magnetic field for spherical harmonic degrees higher than eight. 

The dipole tilt angle in the simulation is also not as small as the one on Saturn. We recently showed that dynamos with moderate Ekman numbers and low convective supercriticalities are able to produce dipole tilts even smaller than Saturn \citep{yadav2022}. However, the model of \citet{yadav2022} lacked many other properties of Saturn, e.g. multiple zonal jets were not present, there were no large scale vortices near the surface, and the magnetic energy in the quadrupolar components was much too low as compared to Saturn. 

Based on the results and discussion above, we believe that the introduction of the SSL is an additional and crucial ingredient needed to promote more Saturn-like features. The presence of an SSL partially separates the outer hydrodynamic layer from the inner dynamo. This is likely needed for promoting multiple off-equatorial zonal jets in outer layers since earlier models without an SSL only produced one equatorial jet outside the magnetic tangent cylinder \citep{duarte2013,dietrich2018,yadav2022}. The SSL also provides a relatively quiescent region where zonal jets, impinging from the outer hydrodynamic layer, can interact with the magnetic field and selectively suppress some of the poloidal field depending on the zonal flow strength and length scale. This, we believe, appears crucial for inducing a selective dip in the 5th spherical harmonic degree component of the poloidal magnetic field. 

Therefore, future improvements will likely need to combine aspects of both studies. For instance, the parameter regime where dynamos with extremely small dipole tilt exist are likely broader at lower Ekman numbers \citep{yadav2022}. There might be a parameter space where the dynamo region in the deep interior itself generates a dipole-dominant solution with very small dipole tilts, and the SSL and zonal jets in the outer layers help further reduce the non-axisymmetric features to minuscule values. With a broader parameter study in future, we hope to answer some of these questions.

\vspace{1.5cm}

\noindent {\bf Acknowledgements}: The work was supported by the NASA Cassini Data Analysis Program (Grant No. 80NSSC21K1128). The computing resources were  provided by the NASA High-End Computing (HEC) Program through the NASA Advanced Supercomputing (NAS) Division at Ames Research Center and the Research Computing, Faculty of Arts \& Sciences, Harvard University. Some of the intermediate simulations were carried out using the computing support provided by the NASA Juno project. RKY thanks Dr.~Ankit Barik for helping to extract the background pressure profile of the simulation.

\vspace{0.5cm}
\noindent {\bf Data Availability}: The simulation input file that can be used to reproduce the results is archived on the Harvard Dataverse: {\tt "https://doi.org/10.7910/DVN/QTHFVS"}. The simulation code used is open access and is available here: {\tt "https://github.com/magic-sph/magic/"}. The large sized output data, which is not practical to be archived on an open access repository, is available upon request to the authors.

\bibliographystyle{abbrvnat}
\bibliography{cited}


\renewcommand\thefigure{\arabic{figure}}
\renewcommand{\figurename}{{\bf Supplementary Figure}}
\setcounter{figure}{0} 
\vspace{1cm}
\vspace{1cm}

\section*{SI Figures}

\begin{figure}[!ht]
\centering
\includegraphics[width=0.9\linewidth]{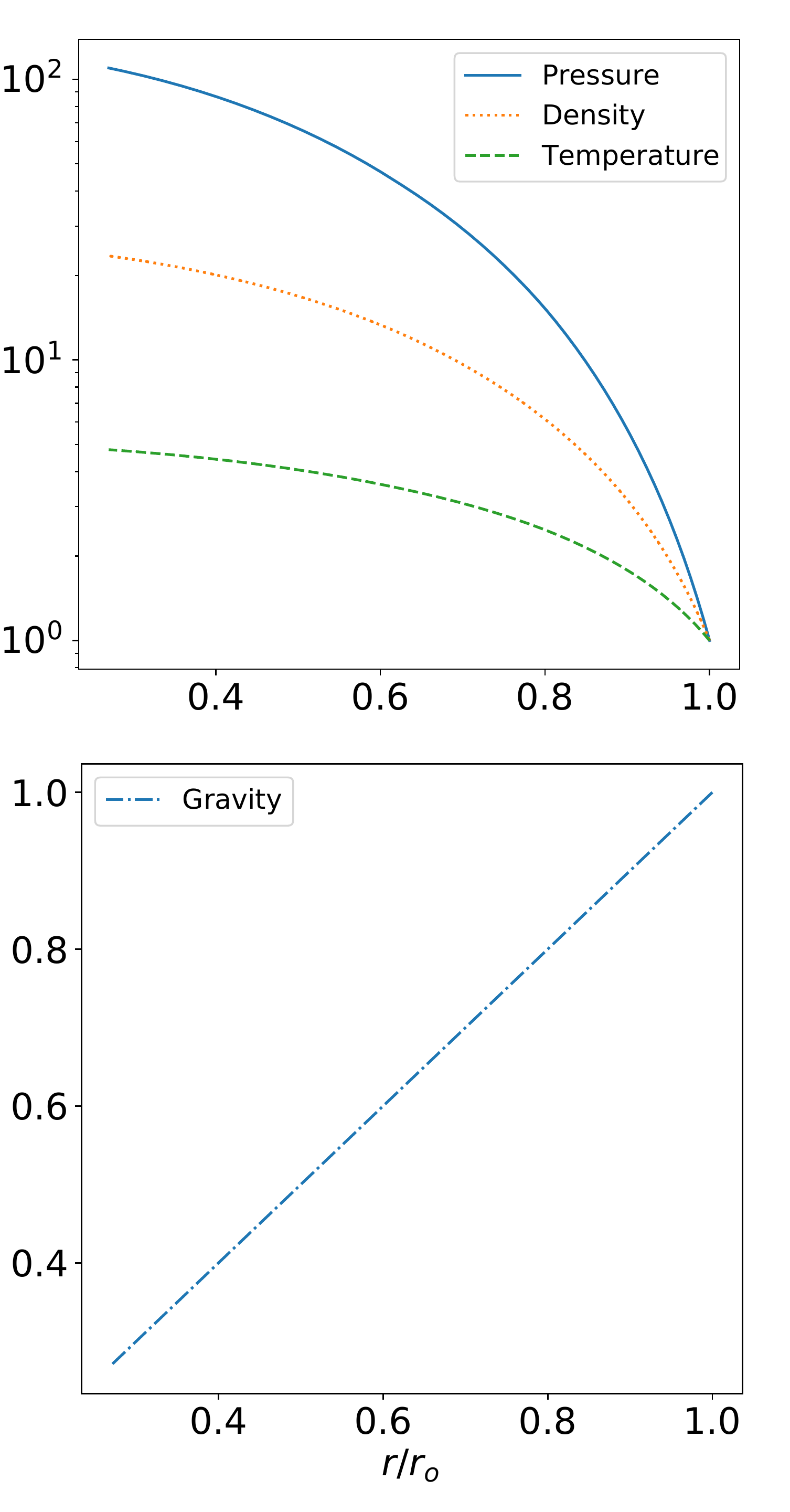}
\caption{Radial variation of the non-dimensional pressure, density, temperature and gravity in the anelastic background state. Note that in the MagIC code, the thermodynamic quantities are normalized by their values at the outer boundary.}
\label{SI_fig1}
\end{figure}

\clearpage

\section*{Legends for the SI animations}

\noindent\textbf{SI Animation 1}: The animations show the evolution of the trajectories of massless particles driven by the local latitudinal-longitudinal flow components (see Fig.~\ref{fig3}). The left panel shows a global overview of the simulation from a northern mid-latitude viewpoint. The right panel views the same animation from the north pole. The trajectories are visualized at radius 0.93$r_o$. The animation, lasting about 3 seconds, shows roughly 12 rotations of the simulation in physical time. The highly non-linear and complex evolution of the zonal jets and vortices is clearly visible in the animation.

\vspace{1cm}

\noindent\textbf{SI Animation 2}: The animation (lasting for about 1m6s) shows the simulation from the viewpoint of a hypothetical orbit around it. The left panel shows the instantaneous magnetic energy as volumetric rendering, along with a meridional slice of the instantaneous local zonal flow as color map (see Fig.~\ref{fig5}a). We have made the smaller magnitude magnetic field either completely transparent or translucent in order to highlight the regions with intense magnetic field. The two rings of azimuthal magnetic energy as well as the central dynamo region are made clearly visible in this way. The right panel shows the instantaneous stream lines of the magnetic field (see Fig.~\ref{fig5}b). The grey spherical surface is located at radius 0.7$r_o$. The rings of magnetic energy visible in the left panel are now visible as magnetic field lines which are mainly wrapped around the sphere.

\end{document}